\documentclass[amsmath,amssymb,aps,prb,twocolumn]{revtex4-2}
\usepackage{graphicx}
\usepackage{dcolumn}
\usepackage{bm}
\usepackage{hyperref}
\usepackage{amsmath}
\usepackage{amsfonts}
\usepackage{amssymb}
\usepackage{natbib}
\usepackage{textcomp,gensymb}
\usepackage{color}

\begin{document}

\title{Surface roughness noise analysis and comprehensive noise effects on
depth-dependent coherence time of NV centers in diamond}
\author{P. Chrostoski$^{\dag,\ast}$}
\author{P. Kehayias$^{\ddagger}$}
\author{D. H. Santamore$^{\ast}$}
\affiliation{$^{\dag}$Sandia National Laboratories, Livermore, California 94550, USA}
\affiliation{$^{\ddagger}$Sandia National Laboratories, Albuquerque, New Mexico 87123, USA}
\affiliation{$^{\ast}$Department of Physics and Engineering Physics, Delaware State
University, Dover, DE 19901, USA}

\begin{abstract}
Noise is a detrimental issue for nitrogen-vacancy (NV) centers in diamond,
causing line broadening and decreasing the coherence time ($T_{2}$).
Following our previous electric and magnetic field noise work, we
investigate noise caused by the diamond surface roughness, which is a source
for charge density fluctuations and incoherent photon scattering. We find
that the varying surface charge density noise source is prevalent throughout
the entire NV dynamical decoupling frequency range, while the photon
scattering noise is almost negligible. Next, we combine the results from
various noise sources to perform comprehensive analyses on $T_{2}$ and how
it varies with NV depth. At a given NV depth of 5 $\mathrm{nm}$ below a
hydrogen- or fluorine-terminated surface, we find that these magnetic
nuclei reduce the NV coherence time the most, followed by the
surface electric field noise sources. The photon scattering and bulk
magnetic field noise effects on $T_{2}$ are weak compared to the varying
charge density, electric dipole, and surface impurity noise. However, with
oxygen surface termination, the surface electric field noise sources are
comparable to the surface magnetic field noise. Our calculated values of  
$T_{2,\text{Hahn}}$ (few $\mu$s to ten $\mu$s) are in good
agreement with the experimental values reported elsewhere. Finally, we
calculate an anticipated signal-to-noise ratio (SNR) for NV AC magnetometry
of external nuclear spins. In our simplified assessment, where some
depth-dependent parameters (e.g.~NV conversion efficiency) are held
constant, we find that shallower NV layers should yield the best SNR, which
is consistent with experimental findings.

\end{abstract}
\date{\today }
\maketitle

\section{Introduction}

Nitrogen-vacancy (NV) centers in diamond are great candidates for quantum
applications, including quantum metrology and sensing, quantum information
processing, and hybrid quantum systems \cite{LRM14,ZSJ13}. NVs can operate
over a wide range of temperatures and environments (including ambient
conditions). They also are useful for sensing magnetic fields, electric
fields, and temperatures at the nanoscale \cite{SCL14,BSB20}. When used to
sense phenomena external to the diamond, placing NVs close to the diamond
surface can improve the signal amplitude and spatial resolution \cite{SSP13,RMS15,HSR15,MHM14,MHK15,FSD16}. However, shallower NVs experience
more surface noise and a faster decoherence rate. This noise broadens the
transition linewidths between NV ground-state sublevels, reduces the
lifetimes of these sublevels, and decreases the overall quantum sensor
performance. Diamond surface noise characteristics are therefore important to
understand, since this informs us of the trade-offs between using shallow or
deep NVs for external sensing.

Electric and magnetic field fluctuations are major NV noise contributors 
\cite{BSS17,BAM14}. Some of the magnetic field noise comes from the nuclear
and electronic spin baths in the bulk \cite{TAK14,OPC12,RDS07,BPB12,BPJ13}.
In addition, the diamond surface also contributes noise due to electron
spins of dangling bonds \cite{OSV09,SZZ80}, terminating surface atoms \cite{JT09,LPM13}, 
adsorption of external molecules \cite{BVW72}, and static
magnetic impurities in thin films \cite{BBK09,PGA09}. Static magnetic
impurities can arise within the bulk naturally or on the surface in thin
films, and have been experimentally observed for both bulk and single-crystal
surfaces \cite{OPC12,KO13,OHB12}.

In our previous work, we showed that the terminating surface atoms
(hydrogen, flourine, and oxygen) often used in experiments can generate more
magnetic field noise than the magnetic impurities ($^{13}$C nuclei) within
the bulk \cite{CBS21}. We also showed that the electric dipole fluctuations
of the diamond surface and different protective surface layers are a large
source of surface electric field noise \cite{CSS18}. Electric field noise is
important, as it causes population decay between the NV $\left\vert
+1\right\rangle $ and $\left\vert -1\right\rangle $ ground-state sublevels \cite{BSS17} and 
decoherence for Autler-Townes dressed states for divacancy defects in 4H-SiC \cite{MBA20}.
Another significant electric field noise source comes from the
diamond surface roughness. A recent experiment found that tri-acid cleaning
and oxygen annealing led to a $4\times $ increase in shallow NV coherence
times \cite{SDS19}. Until now there has been no systematic theoretical study
of electric field noise due to the rough surface in NV quantum sensors.
Taking a comprehensive noise approach and analyzing the noise source
contributions to NV coherence time is crucial to improve quantum sensing
with shallow NVs.

In this paper, we have two objectives: (a) calculate the noise generated by
the rough diamond surface, and (b) provide a comprehensive approach of
determining how noise sources affect coherence time by including all other noise
sources from previous work, and investigate the optimized NV depth.

For rough-surface noise, we study two mechanisms: varying surface charge
distribution fluctuations and photon scattering. A trough created by the
rough surface will trap free surface electrons. The amount of trapped charge
varies among the troughs, causing charge distribution fluctuations that lead
to noise. We model the noise due to the varying charge density using the
Schottky approximation, and generate the noise spectrum by using trapped
charge density statistics \cite{LZM15} (see Sec.\ \ref{sec_SurfCharge}.)

The optical and microwave photons from the NV initialization, readout, and
dynamical decoupling pulses will scatter due to elastic collisions with
atoms in the diamond substrate. The atoms then vibrate, emitting
electromagnetic radiation. To model the photon scattering, we consider
incoherent scattering due to the non-flat rough surface. We use the Green's
function method along with a Gaussian rough surface correlation to determine
the scattering field which will interact with the NV electron spin. With the
scattering field two-time correlation determined, the Wiener-Khinchin
theorem \cite{CYM09} yields the noise power spectrum (see Sec.\ \ref{sec_PhotScatt}).

In Sec.\ \ref{sec_T2comprehensive} we investigate all noise sources present
from previous work \cite{CBS21,CSS18} and this current rough-surface work to
calculate their effect on the inhomogeneous dephasing time $T_{2}^{\ast }$.
We use the Gaussian phase noise approximation \cite{BGA09,SMK96} that
relates the Hahn echo coherence time ($T_{2,\text{Hahn}}$) with the noise
when it has reached a white noise spectrum. We also calculate the surface
noise effect on the longitudinal relaxation time $T_{1}$ by applying Fermi's
Golden Rule for electrical dipole interaction noise \cite{SHP21}, as
described in Sec.\ \ref{sec_T2T1Pred}. After determing $T_{2,\text{Hahn}}$
and $T_{1}$, we can determine $T_{2}^{\ast }$ through the inverse
relationship between $T_{2}^{\ast }$, $T_{2,\text{Hahn}}$, and $T_{1}$.

Recent experiments have seen that NV lifetimes depend on depth \cite{BAM14,RMU15,MSH08}. 
In Sec.\ \ref{sec results_coherence}, we find the depth
dependence of the NV $T_{2,\text{Hahn}}$ and $T_{1}$ lifetimes. We then
calculate the magnetic sensitivity to AC magnetic fields ($\eta$) and the
magnetic field variance from Larmor precession of external nuclei on the
diamond surface ($B_{\text{RMS}}^{2}$), which is the signal strength for NV
NMR spectroscopy. These two factors, $\eta$ and $B_{\text{RMS}}^{2}$,
both contribute to the NV NMR spectroscopy signal-noise-ratio (SNR), and
they have opposite depth dependence: $\eta$ improves with deeper depth
while $B_{\text{RMS}}^{2}$ improves with shallower depth. We then predict
how the SNR for detecting external solid-state nuclei with a shallow NV
varies with depth, as was done experimentally in Ref.~\cite{HKZ22} for $^{11}
$B in hBN.

The work presented here shows that the surface noise plays a major role in
reducing the coherence times of shallow NVs. By understanding and mitigating
this noise, extending the lifetimes for shallow NVs, and determining an
optimal NV depth for external sensing, one could improve the sensitivity of
a wide range of NV sensing applications.

\section{Surface roughness noise models and results\label{sec_model}}

\begin{figure}[ptb]
\centering
\includegraphics[width=\columnwidth]{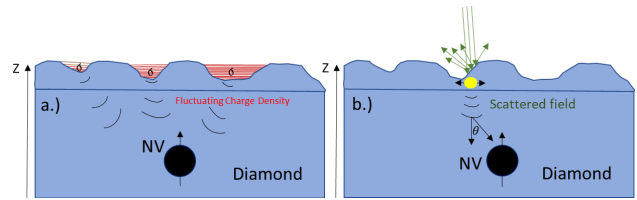}
\caption{Rough surface model for a) valleys creating areas where electrons
can collect and become trapped generating a varying charge density per unit
area due to electrons fluctuating as they repel and try to escape. The
rough surface also gives b) incoherent photon scattering where the scattered
atom will begin to oscillate based on the incident pulse from either an
initialization and readout or dynamical decoupling microwave pulse.}
\label{fig1_RoughSurfaceModel}
\end{figure}

To calculate the rough surface noise, we first identify two noise
mechanisms. The first mechanism is due to rough surface defects (which in
diamond are usually primal $C=C$ sp$^{2}$ bonds \cite{SDC19}) trapping
electrons (see Fig.~\ref{fig1_RoughSurfaceModel}). Depending on the defect
shapes and sizes, some regions have more trapped electrons than others,
causing charge density variations along the surface that leads to
time-dependent fluctuations of the charge density (see Sec.\ \ref{sec_SurfCharge}). 
The second mechanism is the rough surface causing
incoherent scattering of the laser initialization, readout, and dynamical
decoupling microwave pulses, causing photon intensity fluctuations (see
Sec.\ \ref{sec_PhotScatt}). NVs experience the electric fields from
these sources, which enter the NV Hamiltonian as 
\begin{align}
H& =d_{\parallel }E_{z}\left[ S_{z}^{2}-\frac{2}{3}\right] 
\label{EfieldHam} \\
& -d_{\perp }\left[ E_{x}\left( S_{x}S_{y}+S_{y}S_{x}\right) +E_{y}\left(
S_{x}^{2}-S_{y}^{2}\right) \right] ,  \notag
\end{align}
where $d_{\parallel ,\perp }$ are the coupling strengths, $\mathbf{E}$ is
the electric field, and $S_{x,y,z}$ are the NV electron spin operators.

The surface roughness ($Ra$) of Element Six (E6) chemical vapor deposition
(CVD) diamonds commonly used in many experimental setups is $5$ $\mathrm{nm}$
after polishing. E6 can use scaife polishing to smooth the diamond surface
to an $Ra$ as small as $\leq$ $1$ $\mathrm{nm}$ \cite{E610}.

\subsection{Rough surface charge distribution noise \label{sec_SurfCharge}}

To determine the noise spectral density due to the varying charge density,
we use the Schottky approximation to determine the total charge per unit
area (see Ref.~\cite{SDC19} for an approximate solution similar to our
direct derivation), 
\begin{equation}
Q_{d}=\sqrt{2N_{A}\varepsilon\varepsilon_{0}(E_{F}-E_{n})}.
\label{ChargePerUnitArea}
\end{equation}
Here, $\varepsilon$ is the relative permittivity of diamond, $\varepsilon_{0}
$ is the permittivity of free space, $q$ is the electron charge, $N_{A}$ is
the defect concentration, $E_{F}$ is the Fermi energy level, and $E_{n}$ is
the local energy level within the defect region.

Using the total areal charge due to the surface defects, we now consider the
time-dependent charge concentration as the troughs trap surface charges. Xia 
\textit{et al.}~\cite{LZM15} showed that the time-dependent charge
concentration due to trapped electrons is 
\begin{equation}
\rho(t)=qN_{\text{trap}}f(E_{n})\exp\left[ -\frac{\upsilon_{\text{de}}^{2}}{2}t^{2} \right] .   \label{chargeconcdens}
\end{equation}
Here $q$, $N_{\text{trap}},$ and $f(E_{n})$, are the electron charge, number of
trapped electrons, and Fermi-Dirac distribution function at $t=0$,
respectively, and $\upsilon_{\text{de}}=\left( k_{B}T\right) ^{3}/\left(
6h^{3}\upsilon^{2}\right) $. Furthermore, $\upsilon_{\text{de}}$ is the maximum
detrapping rate, $k_{B}$ is Boltzmann's constant, $T$ is temperature, $h$ is
Planck's constant, and $\upsilon$ is the orthogonal vibrational
frequency around the defect. The energy level of the traps (i.e.~the 
Fermi level pinning) is temperature- and time-dependent, and is
represented by $E_{T}=k_{B}T\ln(\upsilon_{\text{de}}t)$. Assuming that $q$ is the
total charge at $t=0$, and Eq.~(\ref{chargeconcdens}) is for the
concentration of charge, $q$ will now be replaced by the total charge per
unit area due to the defects causing roughness, $Q_{d}=\sqrt{
2N_{A}\varepsilon\varepsilon_{o}(E_{F}-E_{n})}$. Plugging in the total
charge per unit area (Eqn.~(\ref{ChargePerUnitArea})) and the Fermi-Dirac
distribution function will give a time-dependent charge density (see
Appendix (\ref{Appdx:density}).

With the time dependence of the charge density, the two-time correlation can
be expressed by an autocorrelation function of the time-dependent charge
density, which can be plugged into the Wiener-Khinchin theorem to determine
the noise spectral density, 
\begin{equation}
S_{CD}(\omega)=\int_{-\infty}^{\infty}\left\langle \delta\rho(t),\delta
\rho(t+\tau)\right\rangle \exp\left[ -i\omega\tau\right] d\tau.
\end{equation}
After applying the Wiener-Khinchin theorem, we get the noise power density,%
\begin{equation}
S_{CD}(\omega)=Q_{d}^{2}N_{\text{trap}}^{2}f(E_{n})^{2}\frac{\exp\left[ -\frac{%
\omega^{2}}{\upsilon_{\text{de}}^{2}}\right] }{\upsilon_{\text{de}}^{2}}\sqrt{2\pi}.
\end{equation}

\subsection{Photon scattering noise \label{sec_PhotScatt}}

As photons from the initialization, readout, and dynamical decoupling pulses
interact with the diamond substrate, atoms within the diamond will have
elastic collisions with incoming photons. The atoms will vibrate, emitting
electromagnetic radiation. The noise generated from the elastic scattering
requires knowing the fields radiated from the scattering. We approach this
by considering the Green's function method applied to the wave equation as
described in the appendix (\ref{Appdx:scatter}). The scattered field due to
the incoherent scattering of the rough surface can be described by the
following electric field, 
\begin{equation}
E(t)=\frac{qa(t)\sin(\theta)}{4\pi\varepsilon_{0}rc^{2}}.
\end{equation}
Here $q$ is the electron charge, $r$ is the radial distance from the
electron, $a(t)$ is the vibrational acceleration of the scattered atom, $%
\theta$ is the scattering angle, and $c$ is the speed of light. We now
create a two-time correlation function and apply the Wiener-Khinchin theorem
to relate to the noise spectral density,
\begin{equation}
S_{PS}(\omega)=\int_{0}^{\infty}\left\langle E(t)E(t+\tau)\right\rangle \exp
\left[ -i\omega\tau\right] d\tau,
\end{equation}
\begin{equation}
S_{PS}(\omega)=E(\theta)^{2}\frac{1}{2\omega_{\text{inc}}(\omega_{\text{inc}}+\omega)}. 
\label{NoisePS}
\end{equation}
Here $E(\theta)^{2}=\frac{q^{2}E_{0}\sin(\theta)}{m_{e}4\pi\varepsilon
_{0}rc^{2}}$ and $m_{e}$ is the electron mass.

To characterize the rough surface causing photon scattering, we need a
correlation function that describes the surface roughness. We describe
the rough surface correlation as follows,
\begin{equation}
C(R)=\frac{1}{\sigma^{2}}\left\langle h(r)h(r+R)\right\rangle ,
\end{equation}
where $h(r)$ is the surface height a distance $r$ away from a smooth
reference plane and $\sigma$ is the root-mean-square height. We use
Gaussian height distributions as they are widely used to describe rough
surfaces. If we consider the rough surface heights that arise from a large
number of random local defects, we can use the central limit theorem to have
the cumulative effect be described using a Gaussian function. To see how the
surface roughness affects the noise due to scattering, we generate the power
spectrum of the rough surface by taking the Fourier transform to get the
rough surface power density,
\begin{equation}
P_{G}(\omega)=\frac{\sigma^{2}\lambda}{4\pi^{3/2}}\exp\left( \frac {%
\lambda^{2}\omega^{2}}{4c^{2}}\right) .
\end{equation}
Here $\omega/c$ is related to the wave number through $k=\omega/c$, and $\lambda$ is the correlation length
(which relates to the surface roughness $Ra$). Now that we have the rough
surface noise power spectrum, we need to combine our noise and rough surface
power densities to generate the actual photon scattering noise power
spectral density seen by the NV spin, giving us the following
\begin{equation}
S_{PS}(\omega)=P_{G}(\omega)E(\theta)^{2}\frac{1}{2\omega_{\text{inc}}(\omega_{\text{inc}}+\omega)},
\end{equation}
where $\omega_{\text{inc}}$ is the incident photon frequency.

\subsection{Results and discussion - rough surface noise\label%
{SurfNoiseResults}}

\begin{figure}[ptb]
\centering
\includegraphics[width=\columnwidth]{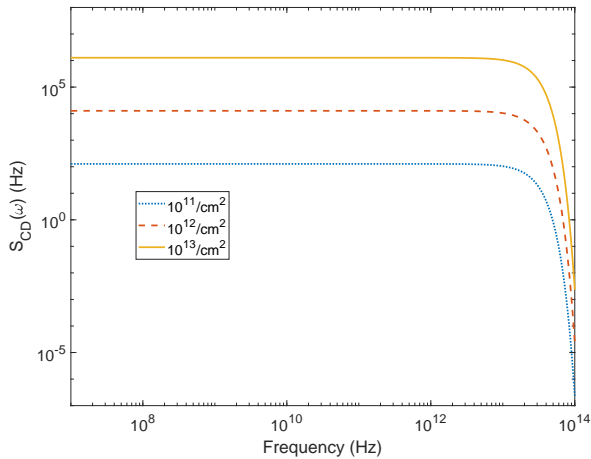}
\caption{Noise power spectra for varying charge distribution noise when
considering a concentration of $10^{13}/$cm$^{2}$ (orange dotted), $10^{12}/$cm$^{2}$ (red dashed), $10^{11}/$cm$^{2}$ (blue) trapped charge area
density.  A $10^{13}/\mathrm{cm}^{2}$   surface charge density is a plausible trapped charge area density due to surface defects \cite{SDC19} before any polishing.}
\label{fig2_CDRoughSurfNoise}
\end{figure}

NV dynamical decoupling experiments often probe the $10^{3}-10^{7}$ $\mathrm{%
Hz}$ operational frequency range, though NVs can also sense higher frequencies
when measuring Rabi frequencies or $T_{1}$ lifetimes. We initially
calculated the noise spectrum from the varying charge density due to the
rough surface for this range, but noticed that the noise very quickly
reaches its maximum, making it nearly constant
throughout the operational frequency range. To see any changes, we expanded
the frequency range (see Fig.~\ref{fig2_CDRoughSurfNoise}). This
almost-constant noise amplitude within the operational frequency range is
associated with the fact that it does not take many electrons getting
trapped to start interacting with each other and vibrate rapidly, generating
noise. The noise amplitude also depends on the square of the electron
density trying to escape from the troughs. The more electrons interacting
within the troughs, the more noise will be generated.

As the surface becomes smooth, the charge density on the surface becomes
more uniform, leading to a large decrease in electrons being trapped in
troughs, reducing the noise. This leads to a smaller noise floor maximum for
a smoother surface, but it does not take many electrons getting trapped in
troughs to start interacting with each other, meaning the noise is
present throughout the entire operational frequency range. Thus, smoothing
out the surface by successive tri-acid cleaning and annealing is effective
in reducing the surface noise \cite{SDS19}.

\begin{figure}[ptb]
\centering
\includegraphics[width=\columnwidth]{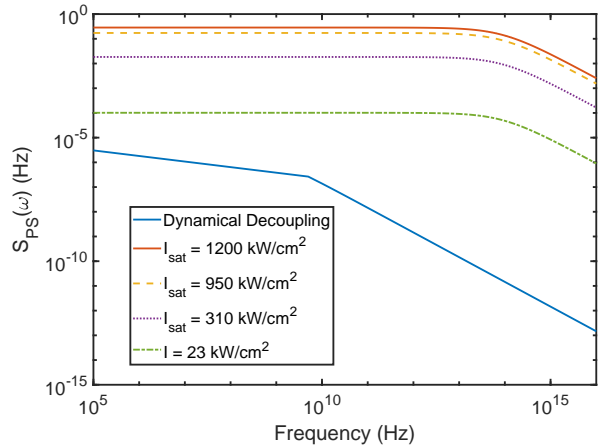}
\caption{Noise power spectra from photon scattering of the initialization and readout laser pulses at the NV optical saturation intensity $I_{\text{sat}} = \frac{h c}{\lambda \sigma \tau}$, where $\lambda = 532$ nm is the pump laser wavelength, $\sigma$ is the NV absorption cross section at 532 nm, and $\tau \approx 10$ ns is the upper-state lifetime. Since there is disagreement in the literature for the value of $\sigma$, here we show the results for $\sigma = 2.4\times10^{-17}$ cm$^{2}$ ($I_{\text{sat}}=1200$ kW/cm$^{2}$, red solid) \cite{SFP18}, $\sigma = 3.1\times10^{-17}$ cm$^{2}$ ($I_{\text{sat}}=950$ kW/cm$^{2}$, orange dashed) \cite{WTH07}, and $\sigma = 9.5\times10^{-17}$ cm$^{2}$ ($I_{\text{sat}}=310$ kW/cm$^{2}$, purple dotted) \cite{CP11}, and the experimental setup of Henshaw et al.~\cite{HKZ22} ($I=23$ kW/cm$^{2}$, green dashed-dotted).}
\label{fig3_IRPulsScattNoise}
\end{figure}

To calculate the noise for the initialization and readout laser pulses and
the dynamical decoupling microwave pulses, the scattering field was set
normal to the NV axis to make the $E(\theta)$ term maximum. This setup
allows us to study the frequency dependence of the noise spectra in addition
to studying its worst-case scenario. Figure \ref{fig3_IRPulsScattNoise}
shows the noise spectrum for the initialization and readout laser pulses. We
considered a $285$ $\mathrm{mW}$ initialization and readout laser
power, $10$ $\mathrm{\mu s}$ pulse duration, and a $40$ $\mathrm{\mu m}$
diameter laser spot size (intensity $I=23$ kW/cm$^2$). \cite{HKZ22}. We also considered several reported NV optical saturation intensities, due to disagreement in the literature \cite{SFP18,WTH07,CP11}. The laser pulses have several orders of magnitude larger noise amplitude than the dynamical decoupling pulses. This is because the laser pulse electric field amplitudes are much
larger than those of the dynamical decoupling microwave pulses (typically
$\sim$10 W microwave power and $\sim$30 ns pulse duration). The peak field
amplitude differences in the pulses leads to the acceleration of the
scattered atoms being much larger for the laser pulses. We also studied
varying the rough surface correlation length from values ranging near the
Van der Waals radius for carbon ($0.17$ $\mathrm{nm}$) to a few millimeters
(a typical diamond size), and saw no change in the noise amplitude. This
indicates that the pulses do not notice the rough surface. Unfortunately,
smoothing the surface does not reduce the noise from photon scattering (like
for the varying charge density). On the other hand, noise during the laser
pulses shouldn't affect the NV lifetimes since they are being optically
pumped. In addition, if the NVs undergo most of their phase accumulation
(and decoherence) in the time between microwave pulses during a dynamical
decoupling AC magnetometry experiment, additional noise during the microwave
pulses also shouldn't matter much (see Section (\ref{sec results
T2Comparison})).

The noise power spectra due to laser pulses and microwave pulses (Fig.~\ref{fig3_IRPulsScattNoise}) and microwave pulses resemble low-pass filters, though with different cutoff frequencies ($\sim$5$\times 10^{13}$ Hz compared to $\sim$5$\times$10$^{9}$ Hz). The laser pulse noise cutoff frequency is beyond the practical NV AC magnetometry frequency range, meaning it's essentially white noise. The microwave pulse noise could be reduced for NV AC magnetometry experiments that surpass this cutoff frequency with an appreciable bias magnetic field. Also note that the microwave pulse noise power spectrum isn’t flat below the cutoff frequency. Finally, NV AC magnetometry experiments should filter out much of this white noise, though the filter pass-bands will still allow some noise through. 

\begin{figure}[ptb]
\centering
\includegraphics[width=\columnwidth]{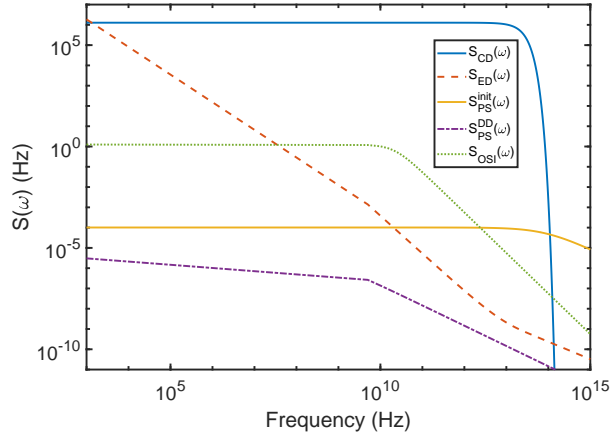}
\caption{Noise power spectra for various surface noise sources considering a 5 nm NV depth, including electric dipole noise considering a bare diamond surface ($S_{ED}(\omega)$, red dashed), varying charge density ($S_{CD}(\omega)$, blue solid), photon scattering from the initialization and readout beams ($S_{PS}^{init}(\omega)$, orange solid), photon scattering from the dynamical decoupling pulses ($S_{PS}^{DD}(\omega)$, purple dashed-dotted), and the magnetic surface impurity noise from oxygen termination ($S_{OSI}(\omega)$ green dotted).}
\label{fig4_SurfNoiseComp}
\end{figure}

We see different power laws for the microwave pulse and laser pulse noise power spectra (Fig.~\ref{fig3_IRPulsScattNoise} and \ref{fig4_SurfNoiseComp}). For the microwave pulse noise, we see a $1/f$ power law at high frequencies. This makes sense as we consider that the photon will scatter off a charged particle within the diamond. This type of scattering will lead to a $1/f$ noise power law \cite{PHH75,PHH80,KVB85}. For the laser pulse noise, we see a $1/f^{2}$ power law at high frequencies. This type of noise can come from a generation-recombination mechanism \cite{MUS19}. This makes sense for the case of the initialization pulse as the NV will absorb the energy and release it through photon emission.

With the noise spectra for the varying charge density and photon scattering determined, we can compare all of the surface noise effects we have studied so far. Fig.~\ref{fig4_SurfNoiseComp} shows the noise contributions from each of the surface noise models we have studied so far. We assume an oxygen-terminated surface without a protective dielectric layer, and a surface roughness that gives a $10^{13}/$cm$^{2}$ trapped charge density. This allows us to study a plausible scenario of the surface noise present in a given experiment before much noise mitigation has been done. For frequencies $< 1$ kHz (not shown), the electric dipole noise is the largest noise source. Continuing to look at the lower frequencies, the electric dipole and varying charge noise density are the largest noise sources. The next largest noise source is the oxygen surface impurity magnetic noise, followed by the photon scattering noise of the initialization and readout pulses when considering the 23 kW/cm$^2$ laser intensity of Henshaw \textit{et al.} \cite{HKZ22}. The smallest surface noise contribution is from the dynamical decoupling microwave pulses, which makes sense when considering the small intensity of the pulses causing a low photon scattering noise floor. These noise profiles will vary based on experimental details, but this gives insight into a scenario for an experiment with limited mitigation done.

\section{Comprehensive noise effects on coherence times and device
sensitivity\label{sec_T2comprehensive}}

\subsection{Coherence time - noise relationship \label{sec_T2NoiseRel}}

With our previous work on electric- and magnetic-field noise \cite{CBS21,CSS18} 
and current work on rough surface noise, we can now take a
comprehensive approach to noise effects on shallow NVs.

\subsubsection{Gaussian phase-noise approximation of $T_{2,\text{Hahn}}$}

Our task is to examine how noise from all these different sources affects the NV
coherence time, $T_{2,\text{Hahn}}$. We assume that the fluctuations will
obey Gaussian statistics \cite{BGA09,SMK96}, i.e.~that the probability
distribution for the stochastic phase of the NV system will follow a Gaussian
\begin{equation}
p(\varphi)=\frac{1}{\sqrt{2\pi\left\langle \varphi^{2}\right\rangle }}%
\exp\left( -\frac{\varphi^{2}}{2\left\langle \varphi^{2}\right\rangle }%
\right) ,   \label{ProbDist}
\end{equation}
where $\varphi$ is the stochastic phase of the NV. To be more specific, we
consider an NV with Bloch vector $\mathbf{M}$ plugged into the Schr\"{o}%
dinger equation, giving $\dot{{\mathbf{M}}\text{ }}=\mathbf{B}\times\mathbf{M%
}$, where $\mathbf{B}$ is the magnetic field. The magnetic field here will have a static part and a stochastic part,
which splits the total phase, $\phi$, into the sum the regular phase, $%
\phi_{0,}$ and the stochastic phase.

In the Gaussian approximation, the only relevant statistical characteristic
is the two-time correlator of the random fluctuation $\left\langle
v(t),v(t+\tau )\right\rangle =W(\left\vert t-\tau\right\vert )$, where $v$
is a random variable. $W(\left\vert t-\tau\right\vert )$ vanishes as $%
\tau\rightarrow \infty$. The integration time will be much larger than the
decay of the correlation function, thus, the central limit theorem becomes
applicable and is independent of the details of the process. The stochastic
phase decay is then given by%
\begin{equation}
e^{-\frac{t}{T_{2}}}=\int p(\varphi)e^{i\varphi}d\varphi=e^{-\frac{1}{2}%
\left\langle \varphi ^{2}\right\rangle }.
\end{equation}
Here $t$ is the integration time and $\left\langle \varphi^{2}\right\rangle $
is the phase variance. The power spectrum of the noise $S(\omega)$ is%
\begin{equation}
S(\omega)=\frac{1}{\pi}\int \nolimits_{0}^{\infty}W(t)\cos(\omega t)dt. 
\label{NoiseDef}
\end{equation}
Combining Eq.~(\ref{ProbDist}) with Eq.~(\ref{NoiseDef}) yields the
following relation%
\begin{equation}
\left\langle \varphi^{2}\right\rangle =4\int \nolimits_{-\infty}^{\infty}%
\frac{\sin^{2}\left( \frac{\omega t}{2}\right) }{\omega^{2}}S(\omega
)d\omega.
\end{equation}
For large $t$, it becomes%
\begin{equation}
\left\langle \varphi^{2}\right\rangle =2\pi tS(0).
\end{equation}
The coherence time relates to the noise at its maximum value $\left(
S(\omega\rightarrow0)\right) $,%
\begin{equation}
T_{2,\text{Hahn}}=\frac{1}{\pi S(0)}.   \label{T2Noise}
\end{equation}

\subsubsection{$T_{2}^{\ast}$, $T_{2,\text{Hahn}}$, and $T_{1}$ noise
\label{sec_T2T1Pred}}

Next, we determine how each noise source contributes to 
$T_{2}^{\ast }$, $T_{2,\text{Hahn}}$, and $T_{1}$ decay. We have modeled each noise
source independently so we can relate the inhomogeneous dephasing time with
the coherence time predicted from each noise source. We follow the $%
T_{2}^{\ast }$\ expression as given in Ref.~\cite{BSB20}, 
\begin{equation}
\frac{1}{T_{2}^{\ast }}\approx \frac{1}{T_{2}^{\text{elect}}}+\frac{1}{T_{2}^{\text{mag}}}+%
\frac{1}{T_{2}^{\text{other}}}+\frac{1}{2T_{1}},  \label{T2Dephase}
\end{equation}%
with%
\begin{equation}
\frac{1}{T_{2}^{\text{elect}}}=\frac{1}{T_{2}^{\text{dip}}}+\frac{1}{T_{2}^{\text{CD}}}+\frac{1}{%
T_{2}^{\text{PS}}},
\end{equation}%
\begin{equation}
\frac{1}{T_{2}^{\text{mag}}}=\frac{1}{T_{2}^{\text{surf}}}+\frac{1}{T_{2}^{\text{bulk}}}.
\end{equation}%
Here $T_{2}^{\text{elect}}$ encompasses the predicted $T_{2,\text{Hahn}}$ for the
electric field noise sources (electric surface dipoles $T_{2}^{\text{dip}}$, photon
scattering $T_{2}^{\text{PS}}$, and varying surface charge density $T_{2}^{\text{CD}}$),
while $T_{2}^{\text{mag}}$ encompasses the predicted $T_{2,\text{Hahn}}$ for the
magnetic field noise sources (electronic surface spin bath $T_{2}^{\text{surf}}$
and nuclear spin bath $T_{2}^{\text{bulk}}$) and the longitudinal relaxation time
of the NV, $T_{1}$. To get a full prediction of $T_{2}^{\ast }$, other
sources that affect the inhomogeneous dephasing time such as strain effects
and other possibly unknown effects will also need to be incorporated.
However, they are out of the scope of this work so we will focus only on how the
various noise profiles we have modeled effect the coherence time. Experiments have also shown that
the Hahn echo pulse technique can yield a $T_{2,\text{Hahn}}$ lifetime much
longer than $T_{2}^{\ast }$ \cite{SSP13,PDC16,SAS12,BSB20,NDH11}.

Determining $T_{1}$ is done by considering the interaction Hamiltonian Eq. (%
\ref{EfieldHam}). If the system is placed close to polar molecules or
dielectric materials, electrical noise from the surface will couple the $%
\left\vert 1\right\rangle $ and $\left\vert -1\right\rangle $ states
together and cause transitions between these states. The rate of this
relaxation can be written using Fermi's Golden Rule \cite{SHP21},%
\begin{equation}
\frac{1}{T_{1}}=\frac{d_{\perp }^{2}}{2}\coth \left( \frac{\beta \omega }{2}%
\right) \int\nolimits_{-\infty }^{\infty }\left\langle \left[ E(t),E(0)%
\right] \right\rangle e^{-i\omega t}dt+\frac{1}{T_{1}^{\text{phon}}}.
\label{T1Equation}
\end{equation}%
Here $\beta =1/k_{B}T$, $\left\langle \left[ E(t),E(0)\right] \right\rangle $
is the thermally-averaged auto-correlation function, and $1/T_{1}^{\text{phon}}$ encompasses
the bulk relaxation rate due to phonons \cite{JAJ12}, including the possibility of surface phonon effects. The Fourier transform in Eqn.~(\ref{T1Equation}) includes the electric field noise from an
electric dipole interaction with a surface dielectric. We have modeled the electric dipole noise due to surface dielectrics previously \cite{CSS18} allowing us to use our previous model to predict the dipole noise effect on $T_{1}$ and get a gauge on $T_{2}^{\ast }$ based on different experimental setups.

\subsubsection{NV relaxation depth dependence due to various noise sources}

Recent experiments have reported how $T_{2,\text{Hahn}}$ depends on the NV depth 
$d$, and the optimal depths for various experimental situations \cite%
{BAM14,RMU15,MSH08,HKZ22}. Equation\ (\ref{T2Noise}) and our work support
the experimental observations: from our previous work on electric dipole
noise \cite{CSS18}, we found that the electric dipole noise had a $1/d^{2}$
dependence. From Eqns.~(\ref{EfieldCD}) and (\ref{NoisePS}) there is
also a $1/d^{2}$ dependence on rough surface noise for both the varying
charge density and photon scattering noises. On the other hand, for the
magnetic noise due to impurity spins at the surface, the magnetic dipole
moment spin-spin interactions give a $1/d^{3}$ dependence \cite{CBS21}.
Finally, for the bulk impurity spins, the spin-spin magnetic dipole moment
interaction depends on the distance between the NV and the $^{13}$C nuclear
spin bath, which is on the order of $0.44$ nm for natural-abundance $^{13}$C
(1.1\%) \cite{HKS16}. In this sense, there is no real depth dependence as
the bulk noise will be felt before any surface effects. Similar to the $T_{2,%
\text{Hahn}}$ depth dependence being driven by the noise, $T_{1}$ also has a
depth dependence as it is affected by the electric dipole noise. We discuss
the depth dependence in details in Sec.\ \ref{sec results_coherence}.

\section{Results and discussion - noise effect on NV lifetimes\label{sec
results_coherence}}

\subsection{Noise effect on T$_{2}$ coherence time\label{sec results
T2Comparison}}

In this section, we calculate the depth-dependent $T_{2,\text{Hahn}}$
coherence time (or decoherence rate) for each noise source using the results
from the previous work \cite{CBS21,CSS18} and the ones obtained in Sec.\ \ref%
{SurfNoiseResults}. For noise spectrum calculations of electric and magnetic
field noise, we refer to our previous work \cite{CBS21,CSS18}. Then, we
compare their effects on $T_{2,\text{Hahn}}$. Table (\ref{table_1}) shows
the coherence time at a depth of $5$ $\mathrm{nm}$ for each noise source. We
did not include the bulk magnetic field noise (e.g.~$^{13}$C and substitutional nitrogen)
since it does not have a clear depth dependence and would obscure the comparison
of other noise effects. We see that the hydrogen- and fluorine-terminated surface impurity magnetic
field noise sources give the shortest $T_{2,\text{Hahn}}$ time. This is due
to the $1/d^{3}$ dependence compared to the $1/d^{2}$ dependence for the
surface electric field noise.

\begin{table}[h]
\begin{center}
\begin{tabular}{|l|l|}
\hline
\textbf{Noise Source} & \textbf{Coherence Time ($\mathrm{\mu s}$) } \\ \hline
Electric dipole (BD) & $3.9$ \\ \hline
Electric dipole (glyc) & $22.5$ \\ \hline
Electric dipole (PC) & $30.8$ \\ \hline
Varying charge density & $4.42$ \\ \hline
Photon scattering & $10^{4}$ \\ \hline
Surface nuclei (F) & $0.004$ \\ \hline
Surface nuclei (H) & $0.08$ \\ \hline
Surface nuclei (O) & $10^{5}$ \\ \hline
\end{tabular}
\end{center}
\caption{Coherence time effects when considering a $5$ $\mathrm{nm}$ depth
from electric dipole noise when considering bare diamond (BD), glycerin
covering layer (glyc), and propylene carbonate covering layer (PC), varying
charge density noise, photon scattering noise from the initialization and
readout pulses with $I_{sat} = 1200$ $\mathrm{kW/cm}^{2}$, and magnetic surface impurities of terminating atoms
fluorine (F), hydrogen (H), and oxygen (O). }
\label{table_1}
\end{table}

The oxygen-terminated surface (which is a widely-used method to get rid of
dangling bonds on the surface) gave a much longer $T_{2,\text{Hahn}}$ time
compared to the electric field noise effects. The depth proportionality
difference of electric field noise and magnetic field noise may explain why
the work by Myers \textit{et al.}~\cite{BAM14} shows the electric field
noise at the surface being comparable to the magnetic field noise. In their
work, they attempt to decompose the noise spectrum into electric field noise
and magnetic field noise since they use the magnetic field variance
proportional to $1/d^{2}$ rather than $1/d^{3}$.

The surface electric dipole contaminations with no protective cover layer
and rough surface with the varying charge density electric field noise gave
the next shortest $T_{2,\text{Hahn}}$ times. When looking at glycerin and
propylene carbonate protective layered electric dipole noise, we see an
order of magnitude increase in the $T_{2,\text{Hahn}}$ time which supports
previous claims that choosing the correct protective layer is important. The
initialization and readout pulse photon scattering noise and
oxygen-terminated surface impurity magnetic field noise give the longest $T_{2,\text{Hahn}}$ times due to their overall addition being masked due to the other stronger noise sources. As expected, the microwave dynamical
decoupling pulses (not included in the table) will also not contribute much to the $T_{2,\text{Hahn}}$ decay rate.

\subsection{$T_{1}$ and $T_{2}^{\ast}$ calculations}

\begin{figure}[ptb]
\centering
\includegraphics[width=\columnwidth]{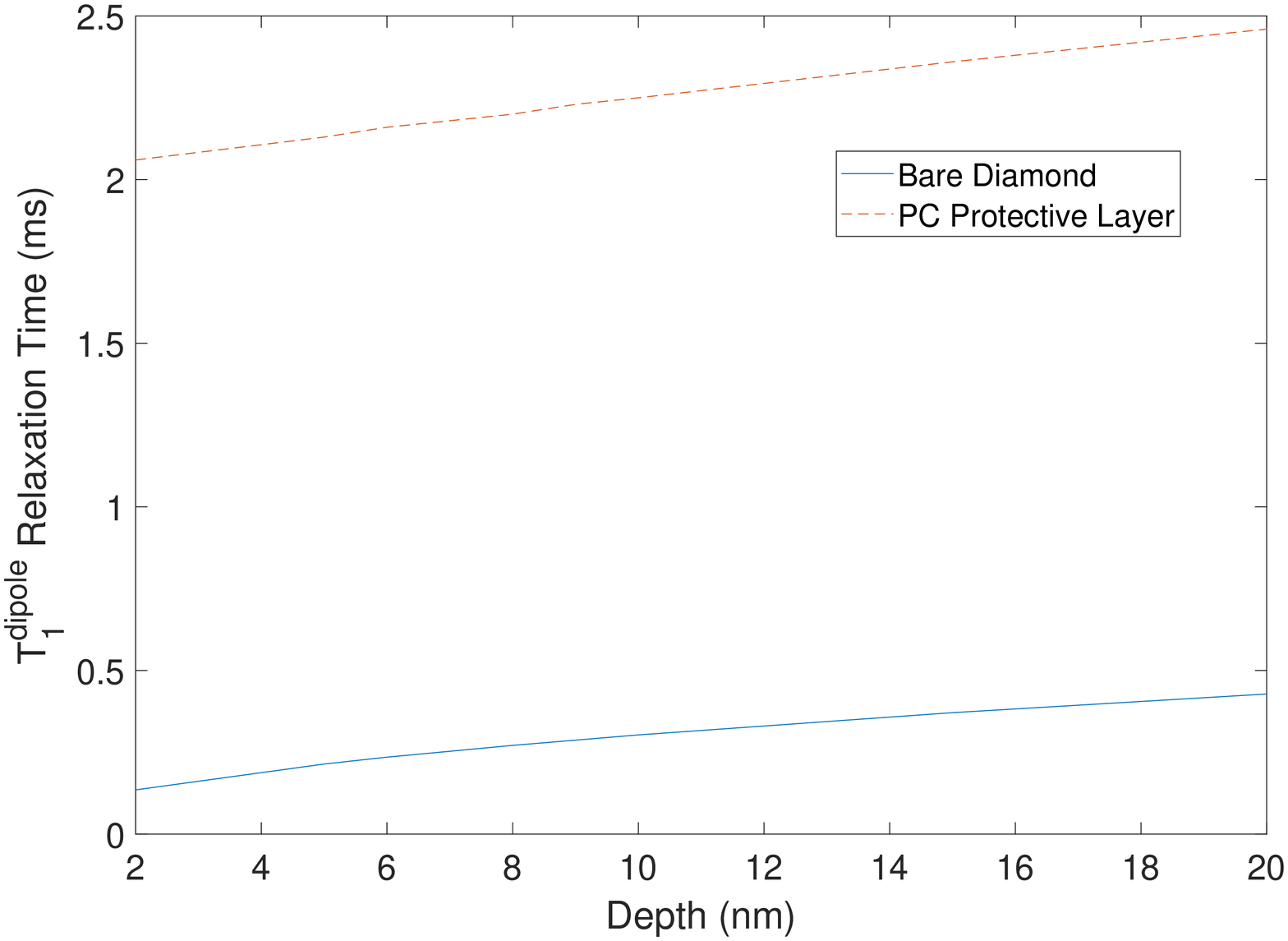}
\caption{$T_{1}^{dipole}$ relaxation times when considering a bare
diamond electric dipole noise (blue line) and a protective layer of
propylene carbonate (PC) (dashed red line). These values are consistent with
the few-ms $T_1$ seen in experiment, though the bulk $T_1$ lifetime is also a few ms.}
\label{fig5_T1VsDepth}
\end{figure}

NV relaxometry experiments have shown that NV room-temperature $T_{1}$ times
are a few milliseconds, depending on the depth \cite{BSS17,BAM14,JAJ12}.
From Eqn.~(\ref{T1Equation}) we see that one part of the $T_{1}$ depth dependence
comes from the electric dipole noise. It should be noted that the $T_{1}^{\text{phon}}$ piece will have a depth dependence of its own. In Fig.~\ref{fig5_T1VsDepth} we calculate $T_{1}$ times due to the surface electric dipole noise at different depths (denoted $T_{1}^{dipole}$) to compare to experimentally-determined values. We look at the cases of a bare diamond surface and a propylene carbonate (PC) protective layer, as this will change the electric dipole interaction strength. Our calculations for $T_{1}$ come out on the order of a fraction of a millisecond for the bare diamond and a few milliseconds with a PC protective layer. These values show good agreement with measured values in the few-millisecond range as seen by Myers \textit{et al.}~\cite{BSS17,BAM14}. Note that the bulk $T_{1}$ lifetime (due to phonons) is also a few milliseconds \cite{JAJ12}.

With the $T_{2,\text{Hahn}}$ decay rates from the different noise sources
and $T_{1}$ determined for different depths, we now use equation (\ref%
{T2Dephase}) to calculate $T_{2}^{\ast}$. Looking at Eq.\ (\ref{T2Dephase}), 
we can see that $T_{1}$ affects $T_{2}^{\ast}$ minimally,
as $T_{1}$ is typically much longer than the $T_{2}^{\ast}$ contributions from the
other noise sources.

\begin{figure}[ptb]
\centering
\includegraphics[width=\columnwidth]{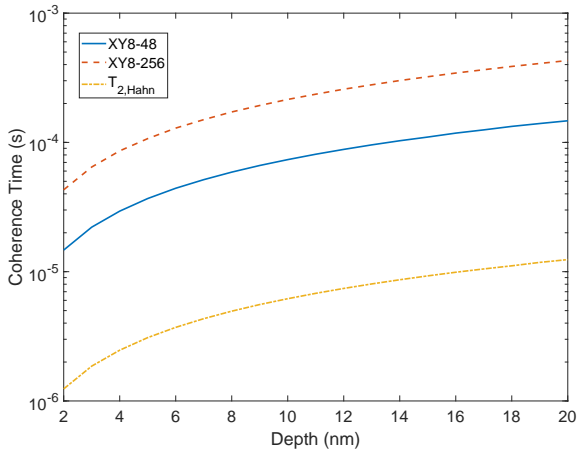} 
\caption{NV coherence times for XY8-48 (solid blue line), XY8-256
(dashed red line), and Hahn echo (dashed-dotted orange line) as a function
of depth. The coherence time improves with depth $d$, as the surface
magnetic field noise is proportional to $1/d^{3}$ and the surface electric
field noise is proportional to $1/d^{2}$.}
\label{fig6_CoherenceTimesVsDepth}
\end{figure}

As mentioned in the previous section, dynamical decoupling pulse sequences can extend the NV coherence times based on the number of $\pi$-pulses $N$. Experimental observations have shown that $T_2$ should improve with $N$ as $T_{2,N}=N^{2/3}T_{2,\text{Hahn}}$, including for shallow NV layers \cite{KJM17}, and we use the same $N^{2/3}$ scaling here. For $N=48$ pulses, we get $T_{2,48}$ values of tens of $\mathrm{\mu s}$ for depths of $2-10$ $\mathrm{nm}$ and $\sim$100 $ \mathrm{\mu s}$ for 20 nm depths, which are in good agreement with experimental observations (see Fig.~\ref{fig6_CoherenceTimesVsDepth}). For $N=256$ pulses, we saw coherence times as long as hundreds of $ \mathrm{\mu s}$. Note that we are only considering the noise effects on $T_{2}$ which we have modeled so far; there may be other effects present (e.g.~strain fluctuations) which can accelerate the decoherence process. Nevertheless, our calculations tell us that these noise sources are playing a big part in decreasing coherence times with surface noise playing a large role. Being able to use methods to mitigate the noise from these different sources will be important to extend these coherence times even further. Some of the suggestions to reduce electric dipole noise are discussed in our previous work \cite{CSS18}.

\subsection{Depth optimization for NV NMR spectroscopy}

In NV NMR spectroscopy of statistically-polarized external nuclei, two
parameters are critical: the magnetic sensitivity and the nuclear magnetic
field amplitude. The sensitivity $\eta$ is the minimum $B_{\text{RMS}}^{2}$
magnetic field noise amplitude that can be measured in a fixed amount of
measurement time. $\eta$ is related to $T_{2}$ in the following way:%
\begin{equation}
\eta\approx\frac{\pi^{2}e\sqrt{T_{2}+T_{R}}}{\gamma_{e}^{2}T_{2}^{2}A\sqrt{%
I_{PL}t_{int}}},
\end{equation}
where $\gamma_{e}$ is the electron gyromagnetic ratio (in MHz/T), $I_{PL}$
is the photon count rate in photons/s, $A$ is the photoluminescence spin
contrast, $t_{int}$ is the readout signal integration time, and $T_{R}$ is
the total readout and initialization time. Here we assume that the noise
floor improves with the square root of the experimental averaging time.
Since the sensitivity contains $T_{2}$, it has a depth dependence, which we
calculate below. For simplicity, we consider the sensitivity of a single NV
rather than an ensemble, since the depth of a single NV is well defined
while the NVs in an ensemble have a range of depths.

In addition, an NV at a depth $d$ below the diamond surface experiences a $%
B_{\text{RMS}}^{2}$ magnetic field from an external semi-infinite
homogeneous layer of nuclei \cite{PDC16}: 
\begin{equation}
B_{\text{RMS}}^{2}=\rho\left( \frac{\mu_{0} h \gamma_{N}}{4\pi}\right)
^{2}\left( \frac{\pi\left( 8-3\sin^{4}\alpha\right) }{128d^{3}}\right) .
\end{equation}
Here, $\mu_{0}$ is the vacuum permeability, $\gamma_{N}$ is the nuclear
gyromagnetic ratio (in MHz/T), $h$ is Planck's constant, $\rho$ is the
nuclear spin density, and $\alpha$ is the angle between the NV axis and the
diamond surface normal vector. For NVs near a [100] diamond surface ($\alpha
\approx54.7\degree$), which is the most common diamond surface cut, this
reduces to 
\begin{equation}
B_{\text{RMS}}^{2}=\rho\left( \frac{\mu_{0} h \gamma_{N}}{4\pi}\right)
^{2}\left( \frac{5\pi}{96d^{3}}\right) .   \label{BRMS}
\end{equation}

\begin{figure}[ptb]
\centering
\includegraphics[width=\columnwidth]{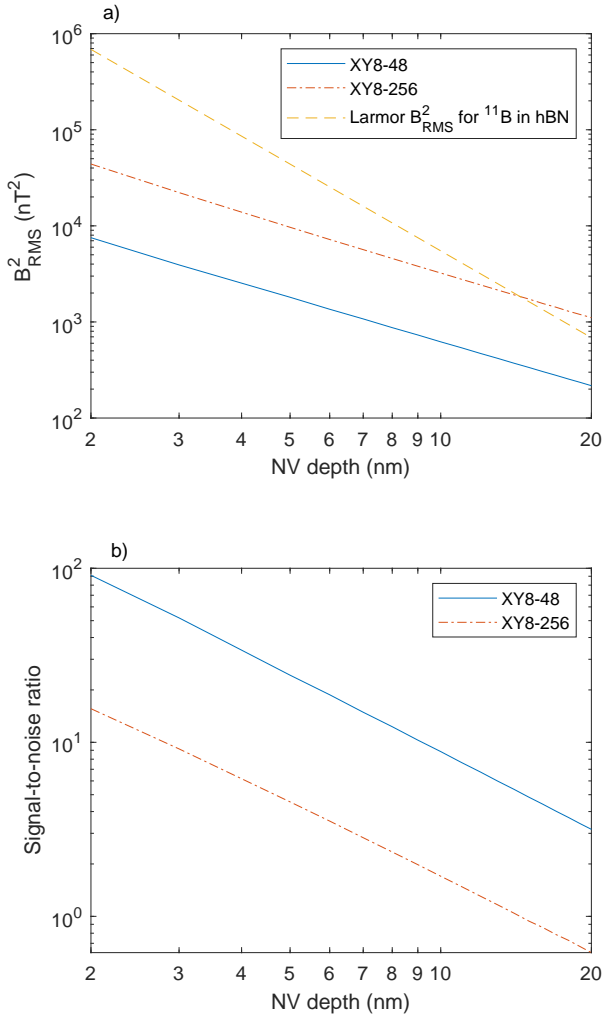}
\caption{(a) $\protect\eta/\protect\sqrt{\text{s}}$ for XY8-48 (solid blue
line), $\protect\eta/\protect\sqrt{\text{s}}$ for XY8-256, (dashed-dotted
red line), and Larmor $B_{\text{RMS}}^{2}$ for $^{11}$B in hBN (dashed
orange line) as a function of depth. $B_{\text{RMS}}^{2}$ improves with
shallower depth like $1/d^{3}$, while the sensitivities get worse (though
more slowly). (b) Signal-to-noise ratio (SNR) for XY8-48 (solid blue line)
and XY8-256 (dashed-dotted red line). The SNR improves for shallower depths,
though in practice other depth-dependent sensitivity factors (e.g.~$I_{PL}$)
move the SNR maximum to a depth greater than zero (e.g.~5.5 nm in Ref.~%
\protect\cite{HKZ22}). }
\label{fig7_SensiPlotAndSNR}
\end{figure}

The signal-to-noise ratio (SNR) between $\eta$ and the $B_{\text{RMS}}^{2}$
is an experimental figure-of-merit: 
\begin{equation}
\mathrm{SNR}= \frac{B_{\text{RMS}}^{2}}{\eta/ \sqrt{\mathrm{s}} } . 
\label{FOM}
\end{equation}
Maximizing the SNR (and minimizing the experiment duration) requires finding
the depth for which an improvement in $B_{\text{RMS}}^{2}$ signal amplitude
is worth the sacrifice in $\eta$. While $\eta$ gets worse with shallower $d$%
, the $B_{\text{RMS}}^{2}$ amplitude improves with shallower $d$, as shown
in Fig.~\ref{fig7_SensiPlotAndSNR} (a). To calculate $\eta$, we used a
typical single-NV fluorescence intensity of $I_{PL} = 2\times10^{5}$
photons/s, $T_{R} = 10~\mu$s, $t_{int} = 2~\mu$s, and $A=0.04$. To calculate 
$B_{\text{RMS}}^{2}$ and compare with Ref.~\cite{HKZ22}, we used $\gamma_{N}
= 13.66$ MHz/T, $\rho= 4.1 \times10^{28}$ spins/m$^{3}$ for $^{11}$B in
hBN, and $\gamma_{e} = 28$ GHz/T.

Figure \ref{fig7_SensiPlotAndSNR} shows how $\eta$ and $B_{\text{RMS}}^{2}$
vary with depth. Here we see that since $B_{\text{RMS}}^{2}$ decreases with
depth more rapidly than $\eta$ does, this means that a shallower depth
should yield a better SNR. This result is consistent with experimental
findings (e.g.~Ref.~\cite{HKZ22, BAM14}), though we leave out additional
depth-dependent phenomena that also affect $\eta$, such as NV conversion
efficiency, photostability, $I_{PL}$, and $A$. In practice, these additional
depth-dependent phenomena can spoil $\eta$ faster than the $1/d^{3}$ $B_{%
\text{RMS}}^{2}$ improvement, meaning the depth for which the NV NMR
spectroscopy SNR is maximized is greater than zero. A more complete analysis
would include these effects, and is out of the scope of this current work.

This NV NMR spectroscopy example (statistically-polarized nuclei in a solid)
does not necessarily generalize to every experimental situation. Applying
this analysis to NV NMR spectroscopy of statistically-polarized nuclei in a
liquid (where the molecular diffusion time depends on NV depth), nuclei with
a $T_{2}^{\ast }$ much shorter than the NV $T_{2}$, and thermally-polarized
external nuclei will likely require modification of the above expressions
for the appropriate depth-dependent SNR \cite{KJM17, GBL18, SDK19}.

Note that we are modeling the worst-case scenarios for the noise sources.
For example, we model the electric dipole noise considering a bare diamond
as in Ref.~\cite{CSS18}. In the actual experiments, there is a hydrocarbon
layer on the diamond surface, and the external nuclei on top of the
hydrocarbon layer \cite{HKZ22}. Depending on the surface dielectric,
the surface electric dipole noise will change and the overall coherence
time could increase, which will change the sensitivity.

\section{Conclusions\label{sec conclusions}}

The studies presented here have been threefold. First we investigated two noise sources due to the rough surface: the varying charge density which comes from electrons on the surface getting trapped within valley defects,
and photons scattering incoherently from the rough surface. Second, we combined all our previous work on electric field noise and magnetic field noise with the rough surface noise and have done comprehensive analyses of the
various noise effects on coherence times, $T_{2,\text{Hahn}}$ and $T_{2}^{\ast}$, and the longitudinal relaxation time, $T_{1}$. Finally, we examined the NV AC magnetic sensing performance with depth and its trade-off
compared to the magnetic field amplitude from external nuclei near the surface.

Our calculations of the varying charge density noise show the white noise present throughout the entire operational frequency range of $10^{3}-10^{7}$ Hz due to the trapped electrons on the surface quickly beginning to fluctuate as they interact with each other. When considering a mostly smooth surface, the noise amplitude decreases considerably but still persists throughout the entire operational frequency range. This agrees with what has been seen in experiment before when using triacid-cleaning to get rid of the surface defects causing valleys \cite{SDS19}. As for the incoherent photon scattering noise due to the rough surface, the noise spectra show that the initialization and readout pulses generate roughly $4-5$ orders of magnitude more noise than the dynamical decoupling pulses. When compared to the other noise sources, the photon scattering is much weaker. When considering a wide range of rough surface correlation lengths we see that the rough surface is not impacting the photon scattering. Methods such as fabricating photonic cavities to reduce the laser power required for initialization or photonic waveguides could help mitigate noise from photon scattering. It should be noted that if optically pumped within a resonant waveguide, the NVs will require less laser power but the photon rate and intensity at the NV will be the same.

We combined this work with our previous work on electric field noise and magnetic field noise \cite{CBS21,CSS18} and examined the effect on coherence time. Our results gave that the surface magnetic impurities,
electric dipole noise, and varying charge density noise play a large role in decreasing coherence times. We next calculated the surface noise effect on the longitudinal relaxation time $T_{1}$, and verified that the result reflects experimentally determined values \cite{BSS17,BAM14} very well. When making our best assumptions to the possible noise sources in an experimental setup similar to Henshaw \textit{et al.}~\cite{HKZ22} and assuming the scaling in the coherence time due to the number of pulses \cite{KJM17}, we saw good agreement for the coherence times for our range of depths when compared to their data.

With the comprehensive analysis on the noise effects on coherence times, we finally optimize the NV depth for magnetometry. Our SNR vs.~depth assessment suggests that shallower NVs are better, a trend similar to what was seen by Henshaw \textit{et al.}~\cite{HKZ22}. Since we only consider noise which we have modeled thus far, the difference in what we see compared to observed by their experiment comes from the fact that their measurements encompass everything that could be decreasing coherence times whereas we only look at one part of the whole picture. Our results show that the noise piece is an important one.

Again, it should be noted that our noise calculations do not account for various experimental efforts (e.g.~tri-acid cleaning and annealing, choosing the correct protective layer, etc.) to reduce noise. However, our results give good predictions to what has been expected from these various noise sources and will aid experimental efforts. Regardless of the numerous experimental improvements to sensitivity, we still see that the noise is playing a major role in the decreasing of coherence times and among all our modeled noise sources, the surface noise is a major culprit. By reducing surface noise, extending lifetimes, and determining an optimal NV depth, one could increase the accuracy and fidelity of a wide range of NV sensing applications.

The predictions presented here can be improved by modeling additional noise sources, like electron spin noise from paramagentic nitrogen defects (P1 centers) and possible electron-electron spin-spin interaction noise from the surface defects. The diamond $^{13}$C isotopic abundance is also an important factor \cite{CBS21} that could modify the effects of some of our current models. Future work decomposing the components of each noise model can help us understand how the noise affects decoherence and relaxation separately, further strengthening the results presented here. Another future direction with these models could be the temperature dependence of the predicted coherence times, as each of the noise models we have described so far may have temperature dependence.

\begin{acknowledgments}
This work is supported by the Advanced Quantum Sensing Center DoD
contract W911NF2020276. We thank Andrew Mounce and Jacob Henshaw of Sandia National Laboratories for helpful discussions and experimental insights. Sandia National Laboratories is a multi-mission laboratory managed and operated by National Technology and Engineering Solutions of Sandia, LLC, a wholly owned subsidiary of Honeywell International, Inc., for the DOE's National Nuclear Security Administration under contract DE-NA0003525. This paper describes objective technical results and analysis. Any subjective views or opinions that might be expressed in the paper do not necessarily represent the views of the U.S. Department of Energy or the United States Government.
\end{acknowledgments}

\appendix

\section{Noise spectrum derivations\label{AppdxF-T}}

\subsection{Varying charge density noise\label{Appdx:density}}

To determine the areal charge density of the rough diamond surface due to
defects, we use the Schottky approximation from the depletion of occupied
states due to surface defects (which in diamond are usually primal $C=C$ sp$%
^{2}$ bonds \cite{SDC19}). The boundary conditions are:

\begin{align}
\rho(x) & =qN_{A} &0<x_{d}\leq x, \\
\rho(x) & =0 &x>x_{d},  \notag
\end{align}
where $q$ is the electron charge, $N_{A}$ is the defect concentration, and $x_{d}$ is
the length of the depletion region. The depletion region is the area where
there is no moving charge so an electric field can be present. With these
boundary conditions, the total charge per unit area within the depletion
region is $Q_{d}=qN_{A}x_{d}$. The next step is to calculate the value for
the length of the depleted region using experimentally available parameters.
Gauss's law states

\begin{equation}
\oint E(x)dA=\frac{Q_{d}}{\varepsilon_{s}},
\end{equation}
where $\varepsilon_{s}$ is the permittivity of the depleted region, $E$ is
the electric field, and $dA$ is the differential area. From our boundary
conditions for our charge density, we can determine how the electric field
will look within our depleted region due to the defects as well as outside
the region,

\begin{align}
E(x) & =-\frac{qN_{A}}{\varepsilon_{s}}(x_{d}-x) &0<x<x_{d}, \\
E(x) & =0 &x\geq x_{d}.  \notag
\end{align}
Here the electric field goes to zero outside the depleted region. This makes
sense as a non-zero field would cause mobile carriers to redistribute
themselves until the field is zero. The maximum possible value for the
electric field is $E_{\max}=-qN_{A}x_{d}/\varepsilon_{s}$.

The electric potential corresponding to the electric field then becomes%
\begin{align}
\phi(x) & =0 &x=0,  \label{EfieldCD} \\
\phi(x) & =\frac{qN_{A}}{2\varepsilon_{s}}\left[ x_{d}^{2}-(x_{d}-x)^{2}%
\right] &0<x<x_{d},  \notag \\
\phi(x) & =\frac{qN_{A}}{2\varepsilon_{s}}x_{d}^{2} &x>x_{d}.  \notag
\end{align}
A boundary condition on the potential is applicable when the density of the
free charge carriers is very high and the thickness of the charge layer is
very thin, thus, a potential difference between them is orders of magnitude
smaller than the potential difference within the defect despite the total
amount of charge being the same. This is a key piece of the Schottky
approximation and works well when determining the energy bands at the
diamond surface. The total potential difference within the depletion region
is the Fermi energy ($E_{F}$), that is further reduced or increased by the
local energy level, $E_{n}$, in the depletion region. The boundary condition
gives
\begin{equation}
E_{F}-E_{n}=-q\phi(x=0)=\frac{q^{2}N_{A}}{2\varepsilon_{s}}x_{d}^{2}
\end{equation}
and the length of the depleted region becomes
\begin{equation}
x_{d}=\sqrt{\frac{2\varepsilon_{s}(E_{F}-E_{n})}{q^{2}N_{A}}},
\end{equation}
With the relation between the maximum electric field, we obtain the total
charge per unit area which is similar to the approximation found in
Ref.~\cite{SDC19}, 
\begin{equation}
Q_{d}=\sqrt{2N_{A}\varepsilon\varepsilon_{o}(E_{F}-E_{n})}.
\end{equation}
where $\varepsilon$ is the relative permittivity of diamond and $\varepsilon
_{o}$ is the permittivity of free space.

Using the total areal charge due to the surface defects, we now consider the
time-dependent charge distribution as the defects trap mobile charges at the
surface. Xia \textit{et al.}~\cite{LZM15} showed that the time-dependent
charge concentration due to trapped electrons is
\begin{equation}
\rho(t)=qN_{\text{trap}}f(E_{n})\exp\left[ -\int
\nolimits_{0}^{t}P_{\text{de}}dt'\right] .
\end{equation}
Here $q$, $N_{\text{trap}},$ and $f(E_{n})$, are the electron charge, trap density occupied,
and the Fermi-Dirac distribution function at $t=0$, respectively, and $%
P_{\text{de}} $ is the probability of an electron being detrapped. The probability
of an electron to be detrapped can be written as a Boltzmann rate,%
\begin{equation}
P_{\text{de}}  =\upsilon_{\text{de}}\exp\left[ -\frac{E_{T}}{k_{B}T}\right] ,
\text{where }\upsilon_{\text{de}} =\frac{\left( k_{B}T\right) ^{3}}{%
6h^{3}\upsilon^{2}}.  
\end{equation}
Here $\upsilon_{\text{de}}$ is the maximum detrapping rate, $k_{B}$ is
Boltzmann's constant, $T$ is temperature, $h$ is Planck's constant, and $%
\upsilon$ is the orthogonal vibrational frequency around the defect. The
energy level of the traps is temperature- and time-dependent and is
represented by $E_{T}=k_{B}T\ln(\upsilon_{\text{de}}t)$. Plugging in the definition
of $E_{T}$ and $\upsilon_{\text{de}}$ into $P_{\text{de}}$ makes determining the probability of
detrapping a straight forward process and gives the the time-dependent
charge concentration as%
\begin{equation}
\rho(t)=qN_{\text{trap}}f(E_{n})\exp\left[ -\frac{\upsilon_{\text{de}}^{2}}{2}t^{2}%
\right] ,
\end{equation}%
\begin{equation*}
\text{where }f(E_{n})=\left[ 1+\exp\left( \frac{E_{F}-E_{n}}{k_{B}T}\right) \right]
^{-1}.
\end{equation*}
Assuming that $q$ is the total charge at $t=0$, $q$ will now be replaced by
the total charge per unit area due to the defects causing roughness, $Q_{d}=%
\sqrt{2N_{A}\varepsilon\varepsilon_{o}(E_{F}-E_{n})}$. Plugging in the
total charge per unit area from before we get the following time-dependent charge density,%
\begin{equation}
\rho(t)=Q_{d}N_{\text{trap}}f(E_{n})\exp\left[ -\frac{\upsilon_{\text{de}}^{2}}{2}t^{2}%
\right] .
\end{equation}

With the time dependence of the charge density, the two-time correlation can
be expressed by an auto-correlation function of the time-dependent charge
density%
\begin{equation}
\left\langle \delta \rho (t),\delta \rho (t+\tau )\right\rangle =\Lambda 
\frac{\exp \left[ \frac{\upsilon _{\text{de}}^{2}\tau ^{2}}{4}\right] \sqrt{\pi }}{%
\upsilon _{\text{de}}},
\end{equation}%
\begin{equation*}
\text{where }\Lambda =Q_{d}^{2}N_{\text{trap}}^{2}f(E_{n})^{2}\exp \left[ -\frac{%
\upsilon _{\text{de}}^{2}}{2}\tau ^{2}\right] .
\end{equation*}%
The auto-correlation function can then be plugged into the Wiener-Khinchin
theorem to relate to the noise spectral density,%
\begin{equation*}
S_{CD}(\omega )=\int_{-\infty }^{\infty }\left\langle \delta \rho
(t),\delta \rho (t+\tau )\right\rangle \exp \left[ -i\omega \tau \right]
d\tau .
\end{equation*}%
After applying the Wiener-Khinchin theorem, we get the noise power density,%
\begin{equation}
S_{CD}(\omega )=Q_{d}^{2}N_{\text{trap}}^{2}f(E_{n})^{2}\frac{\exp \left[ -\frac{%
\omega ^{2}}{\upsilon _{\text{de}}^{2}}\right] }{\upsilon _{\text{de}}^{2}}\sqrt{2\pi }.
\end{equation}

\subsection{Incoherent photon scattering\label{Appdx:scatter}}

\subsubsection{Green's function method}

As photons from the initialization, readout, and Hahn echo pulses interact
with the diamond substrate, atoms within the diamond will have elastic
collisions with incoming photons. The atoms will vibrate emitting
electromagnetic radiation. This requires knowing the fields radiated from
the scattering. We approach this by considering the Green's function method
applied to the wave equation. The potentials from the pump laser pulse can
usually be considered from the Lorenz gauge or Gaussian. Either way the
wave equation will take the following form,%
\begin{equation}
\nabla^{2}\Phi-\frac{1}{c^{2}}\frac{\partial^{2}\Phi}{\partial t^{2}}=-4\pi.
\end{equation}
Here $\Phi$ can be either a scalar potential or a component of the
potential. This gives the corresponding Green's function equation,%
\begin{equation}
\nabla^{2}G(\vec{x},t;\vec{x}',t')-\frac{1}{c^{2}}\frac{\partial^{2}G}{\partial t^{2}}=-4\pi\delta (%
\vec{x}-\vec{x}')\delta(t-t'),
\end{equation}
where the source is now an event located at $\vec{x}=%
\vec{x}'$ happening at $t=t'$. Performing a
Fourier transform and considering the spherical symmetry and the properties
of the delta function give a solution for the Green's function Fourier
transform,%
\begin{equation}
G(\vec{x},\omega;\vec{x}',t')=%
\frac{1}{\sqrt{2\pi}R}\left( Ae^{ikR}+Be^{-ikR}\right) e^{-i\omega
t'}.
\end{equation}
Doing the inverse transform to get us back our original Green's function,%
\begin{equation}
G(\vec{x},t;\vec{x}',t')=\frac {1%
}{\sqrt{2\pi}}\int_{-\infty}^{\infty}G(\vec{x},\omega ;%
\vec{x}',t')e^{-i\omega t}d\omega,
\end{equation}%
\begin{equation}
G(\vec{x},t;\vec{x}',t')=A%
\delta(t'-(t-\frac{R}{c}))+B\delta(t'-(t+\frac{R}{c})).
\end{equation}
The second term is usually rejected as it predicts a response to an event
occurring in the future, so here we shall do the same. The time $t-R/c$ here
is normally referred to as the retarded time $t_{ret}$. With our Green's
function we can solve our wave equation and determine the potential,%
\begin{equation}
\Phi(\vec{x},t)=\frac{1}{c}\int\frac{\vec{j}(%
\vec{x}',t')}{R}\delta(t'-t_{ret})dt'd^{3}x',
\end{equation}%
\begin{equation}
\Phi(\vec{x},t)=\frac{1}{c}\int\frac{\vec{j}(%
\vec{x}',t_{ret})}{R}d^{3}x'. \label{PotentialSolution}
\end{equation}
It can be seen relatively easily that this represents a static
potential, but for our case we want the field from a charge that is
accelerating due to it interacting with the photon.

\subsubsection{Scattered field solution}

Our source of the scattered field is going to be an atom, more specifically
a charge, accelerating due to the elastic collision with the incoming
photon. This will allow us to rewrite Eq.\ (\ref{PotentialSolution}) as the
following,%
\begin{equation}
\Phi(\vec{x},t)=\frac{1}{c}\int\frac{q\vec{v}\delta(%
\vec{x}'-\vec{r}(t'))}{R}%
\delta(t'-t_{ret})dt'd^{3}x'.
\end{equation}
Here $\vec{v}$ is the velocity of the oscillating charge and $%
\vec{r}(t')$ is the position changing over time. 
Doing the integral over the spatial coordinates we get,%
\begin{equation}
\Phi(\vec{x},t)=\frac{1}{c}\int q\vec{v}\frac {%
\delta(t'+R(t')/c-t)}{R(t')}dt',
\end{equation}
where $R(t')=\left\vert \vec{x}'-%
\vec{r}(t')\right\vert $. With what we have here, we
cannot do a straight integration of $t'$. We will have to
reexpress the delta function to do this integral to be the delta
function of a function which has the form,%
\begin{equation}
\delta(f(t'))=\sum\frac{1}{\left\vert
f'(t_{i}')\right\vert }\delta(t'-t_{i}').
\end{equation}
where $f(t_{i}')=0$. Taking the derivative and using the
definition of the velocity, we will get,%
\begin{equation}
f'(t_{i}')=1+\frac{1}{c}\frac{dR}{dt'}=1-%
\frac{1}{c}\frac{(\vec{x}-\vec{r}(t'))}{%
\left\vert \vec{x}-\vec{r}(t')\right\vert }%
\cdot\frac{d}{dt'}(\vec{x}-\vec{r}%
(t')),
\end{equation}%
\begin{equation}
f'(t_{i}')=1-\frac{\vec{v}\cdot (%
\vec{x}-\vec{r}(t'))}{c\left\vert 
\vec{x}-\vec{r}(t')\right\vert }=1-\frac{%
\vec{v}\cdot\vec{R}}{cR}.
\end{equation}
This function is zero for $t=t_{ret}$, so evaluating the integrals we
get
\begin{equation}
\Phi(\vec{x},t)=\frac{q\vec{v}}{R(1-\frac {%
\vec{v}\cdot\vec{R}}{cR})}.
\end{equation}
This is the Leinhard-Wiechart potential, and from here we can start
determining the scattering field.

The electric field from electromagnetic radiation is proportional to the
magnetic field as $\vec{\textbf{B}}=\vec{\textbf{E}}c$. From this, the field which will have the largest
interaction with the NV center electron spin will be the electric field.
Thus, we only consider the electric field emitted from the Rayleigh
scattering within the diamond lattice. We can write the electric field as
follows:%
\begin{equation}
\vec{E}(\vec{x},t)=-\vec{\nabla}V-\frac{1}{c%
}\frac{\partial\vec{\Phi}}{\partial t},
\end{equation}
where the potential $V$ looks like $\Phi(\vec{x},t)$ without the
velocity factor. The potentials here are in terms of $\vec{x}$
and $t_{ret}$, so the partial derivatives will be a bit different, but we can
put the origin at the spontaneous position of the oscillating charge, $R=r$,
allowing us to simplify things. Rewriting the electric field to fit our
potentials in terms of $\vec{x}$ and $t_{ret}$ we get
\begin{equation}
\vec{E}(\vec{x},t_{ret})=\vec{\nabla}V\cdot
d\vec{x}-\frac{\partial\vec{\Phi}}{\partial t_{ret}}%
\frac{dr}{c}+\frac{\partial\vec{\Phi}}{\partial t_{ret}}dt.
\end{equation}
Looking at the first term of the electric field we get,%
\begin{equation}
\vec{\nabla}V=\vec{\nabla}\frac{q}{r}=-\frac{q}{r^{2}}%
\vec{\nabla}r,
\end{equation}
where%
\begin{align}
\vec{\nabla}r & = \\
& \frac{\partial}{\partial r}\left( r-\frac{\vec{r}\cdot%
\vec{v}}{c}\right) \widehat{r}+\frac{\widehat{\theta}}{r}\frac{%
\partial}{\partial\theta}\left( r-\frac{\vec{r}\cdot%
\vec{v}}{c}\right)  \notag \\
& +\frac{\widehat{\phi}}{r\sin\theta}\frac{\partial}{\partial\phi}\left( r-%
\frac{\vec{r}\cdot\vec{v}}{c}\right) .  \notag
\end{align}
Choosing our axes with polar axis along the instantaneous direction of
the velocity, we get $\vec{r}\cdot\vec{v}%
=rv\cos\theta$ giving us,%
\begin{equation}
\vec{\nabla}r=\left( 1-\frac{v}{c}\cos\theta\right) \widehat{r}+%
\frac{\widehat{\theta}}{r}\left( r\frac{v}{c}\sin\theta\right) .
\end{equation}
We can consider here the non-relativistic limit such that $v/c\ll1$,
simplifying the first term in our electric field to be $\vec{%
\nabla }r=\widehat{r}$. To get the entire electric field, we will need the $%
\partial r/\partial t$ term, which will
simply give $\partial r/\partial t=-\left( \vec{r}\cdot%
\vec{a}\right) /c$. Putting everything together, we have%
\begin{equation}
\vec{E}=\frac{q}{r^{2}}\widehat{r}\left( 1+\frac{\left( 
\vec{r}\cdot\vec{a}\right) }{c}\right) -\frac {q%
\vec{a}}{cr}+\frac{q\vec{v}}{cr^{2}}\left( r-\frac{%
\left( \vec{r}\cdot\vec{a}\right) }{c}\right) ,
\end{equation}%
\begin{equation}
\vec{E}=\frac{q}{r^{2}}\widehat{r}\left( 1+\frac{\left( 
\vec{r}\cdot\vec{a}\right) }{c}\right) -\frac {q%
\vec{a}}{cr}.
\end{equation}
The first term is the usual point charge term, but the other two terms
are the radiated field.

You can also look at the trigonometry and see that the parallel component is
the piece that resembles a static charge where as the emitted field will
then be the perpendicular piece. Doing the full vector analysis, we will get the radiated
component of the electric field, 
\begin{equation}
E(t)=\frac{qa(t)\sin(\theta)}{4\pi\varepsilon_{0}rc^{2}}.
\end{equation}
Here $q$ is the electron charge, $\varepsilon_{0}$ is the
permittivity of free space, $r$ is the radial distance from the electron, $%
a(t)$ is the vibrational acceleration of the charge, $\theta$ is the
scattering angle, and $c$ is the speed of light. We neglect the static field
term as physically its interaction with the NV center electron will be much
smaller that of the emitted field from the oscillating charge, so it will not
add to the noise.

\subsubsection{Rayleigh scattering noise density}

Now that we have the emitted field due to the scattering, we need to
determine the acceleration of the oscillating charge. We can do this by
considering the position of an electron bound to an atom in an applied
oscillating electric field,%
\begin{equation}
x_{e}(t)=\frac{qE_{0}}{m_{e}(\omega_{0}^{2}-\omega_{\text{inc}}^{2})}\exp
[-i\omega_{\text{inc}}t].
\end{equation}
Here $E_{0}$ is the amplitude of the electric field, $m_{e}$ is the mass of
the electron, $\omega_{0}$ is the resonant frequency of the diamond, and $%
\omega_{\text{inc}}$ is the frequency of the oscillating field (incident
light-wave). For both pulses, $\omega_{\text{inc}}\gg\omega_{0}$, and from this we
get the acceleration of the oscillating charge will be%
\begin{equation}
a(t)=-\frac{qE_{0}}{m_{e}}\exp[-i\omega_{\text{inc}}t].
\end{equation}
It is important to note that the Hahn echo pulse could be near this resonant
(microwave) frequency, possibly blowing up the above expression, but if we calculate 
$E_{0}$ from experimental values we will get an extremely small amplitude
counteracting this large value from $1/(\omega_{0}^{2}-\omega_{\text{inc}}^{2})$
leading to a small acceleration. From this we will continue with what is
derived here, as the method will not change.

Now that we have a time-dependent electric field, we can determine the
two-time correlation function. One constraint to place is that at $t<0$ the
correlation will go to zero when considering causality. This can also be
done as to determine our potentials the Green's function required a similar
causality. This will allow us to shift the integration limits for the
correlation function integration as well as deal with any convergence issues
of the exponential function. The two-time correlation function is then
defined as%
\begin{equation}
\left\langle E(t)E(t+\tau )\right\rangle =E(\theta )\int\limits_{0}^{\infty
}e^{[-i\omega _{\text{inc}}t]}e^{[-i\omega _{\text{inc}}(t+\tau )]}d\tau ,
\end{equation}%
\begin{equation*}
\text{where }E(\theta )=\frac{q^{2}E_{0}\sin (\theta )}{m_{e}4\pi,
\varepsilon _{0}rc^{2}}
\end{equation*}%
\begin{equation}
\left\langle E(t)E(t+\tau )\right\rangle =E(\theta )^{2}\frac{\exp [-i\omega
_{\text{inc}}\tau ]}{2\omega _{\text{inc}}}.
\end{equation}%
With the two-time correlation function, we can apply the Wiener-Khinchin
theorem and the same causality constraint as before to shift the integration
limits and get rid of any convergence issues to get the noise spectral
density,%
\begin{equation}
S_{PS}(\omega )=\int_{0}^{\infty }\left\langle E(t)E(t+\tau )\right\rangle
\exp \left[ -i\omega \tau \right] d\tau ,
\end{equation}%
\begin{equation}
S_{PS}(\omega )=E(\theta )^{2}\frac{1}{2\omega _{\text{inc}}(\omega _{\text{inc}}+\omega )}%
.
\end{equation}

\subsubsection{Rough surface scattering density}

To characterize rough surface causing photon scattering, we need a
correlation function to describe the surface roughness. In this
sense, we will describe the rough surface correlation as
\begin{equation}
C(R)=\frac{1}{\sigma^{2}}\left\langle h(r)h(r+R)\right\rangle ,
\end{equation}
where $h(r)$ is the surface height a distance $r$ away from a
smooth reference plane and $\sigma$ is the root-mean-square height. Here we use a Gaussian correlation function of the rough surface. It should be noted that the correlation function can also be described with an exponential, which data is often fit to.

To see how the surface roughness will effect the noise due to scattering, we
need to generate the power spectrum of the rough surface by taking the
Fourier transform
\begin{equation}
P(k)=\frac{\sigma^{2}}{\left( 2\pi\right) ^{2}}\int_{-\infty}^{\infty
}C(R)\exp\left( ik\cdot R\right) dR.
\end{equation}
Here $k$ is the wave number and is related to the to the frequency of the
scattered wave by $k=\omega/c$. Plugging in the Gaussian correlation
functions and doing the integral, we get following power density
\begin{equation}
P_{G}(\omega)=\frac{\sigma^{2}\lambda}{4\pi^{2/3}}\exp\left( \frac {%
\lambda^{2}\omega^{2}}{4c^{2}}\right) ,
\end{equation}
where $\lambda$ is the correlation length. Now that we have the rough
surface, we need to combine our noise and rough surface power densities to
generate the actual photon scattering noise power spectral density seen by
the NV center spin giving us%
\begin{equation}
S_{PS}(\omega)=P_{G}(\omega)E(\theta)^{2}\frac{1}{2\omega_{\text{inc}}(\omega_{\text{inc}}+\omega)}.
\end{equation}

\end{document}